\documentclass[twocolumn, prl]{revtex4}
\usepackage{graphicx}
\usepackage{epsfig}
\usepackage{color}
\usepackage{amsmath}
\begin{document}
\title{Contact matrix in dilute quantum systems}
\author{ Shao-Liang Zhang$^*$, Mingyuan He$^*$, and  Qi Zhou}
\affiliation{Department of Physics, The Chinese University of Hong Kong, Shatin, New Territories, HK}
\date{\today}
\begin{abstract}
Contact has been well established as an important quantity to govern dilute quantum systems, in which the pairwise correlation at short distance traces a broad range of thermodynamic properties.  So far,  studies have been focusing on contact in individual angular momentum channels. Here, we point out that, to have a complete description of the pairwise correlation in a general dilute quantum systems,  contact should be defined as a matrix. Whereas the diagonal terms of such matrix include contact of all partial wave scatterings, the off-diagonal terms, which elude previous studies in the literature, characterise the coherence of the asymptotic pairwise wavefunction in the angular momentum space and determine important thermodynamic quantities including the momentum distribution. Contact matrix allows physicists to access unexplored connections between short-range correlations and macroscopic quantum phenomena. As an example, we show the direct connection between contact matrix and order parameters of a superfluid with mixed partial waves. 

\end{abstract}

\maketitle

Since S. Tan first invented the concept of contact in quantum dilute systems in 2005, the study of contact and the universal thermodynamic relations have become a fundamentally important topic in ultracold atom physics, and have also influenced considerably related fields \cite{Tan1,Tan2,Tan3}. Due to the length scale separation that the average interparticle distance $k_F^{-1}$, where $k_F$ is the Fermi momentum, is much larger than the range of interaction $r_0$ in a dilute quantum system, the correlation between a pair of particles at short distance determines a wide range of thermodynamic quantities, and allows physicists to establish deep connections among very different physical quantities. Within a decade, both theoretical and experimental efforts have made significant progresses towards unveiling thermodynamic relations that are universal regardless of the details of microscopic physics \cite{V1,V2,V3,V4,Jin1,Jin2,Jin3,Vale,T1,T2,T3,T4,T5,Zhou}. 

Contact was originally invented for systems with a delta-function interaction $U({\bf r}_i-{\bf r}_j)\sim \delta ({\bf r}_i-{\bf r}_j)$, where ${\bf r}_i$ is the real space coordinate of the $i$th particle in the system \cite{Tan1,Tan2,Tan3}. For such modelling potential, only $s$-wave scattering exists, and the $s$-wave contact alone is sufficient to characterise thermodynamics of the many-body system. Recently, it was realised that contact also exist for $p$-wave scattering\cite{P1,P2,Zhou1}, and an experiment has probed two $p$-wave contact \cite{P3}. In particular, in reference \cite{Zhou1}, we pointed out that contact can be defined for a general short-range interaction including any partial wave scatterings. Putting all these contact together, one gets contact spectrum, which enables universal relations beyond thermodynamics and has powerful applications in atomic quantum Hall states. Later,  a work obtained consistent result for the momentum distribution determined by $d$-wave contact \cite{P4}. Another work also found out in a system with both strong $s$ and $p$-wave interactions, the thermodynamics replies on multiple contact\cite{Cui}. 

In this Letter, we point out that, to completely describe the pairwise correlation at short distance in a generic dilute many-body system, contact should be defined as a matrix $C_{\alpha\beta}$, where $\alpha$ is the short-hand notation of the quantum numbers of angular momentum. For instance,  $\alpha=(l,m)$ and $\alpha=l$ in three and two dimensions, respectively. The diagonal terms, $C_{\alpha\alpha}$, is the contact studied in the literature for an individual partial wave scattering. The importance of the off-diagonal terms, $C_{\alpha\neq\beta}$, are summarised as follows. First,  $C_{\alpha\neq\beta}$ have the same origin as diagonal terms. When the distance between two  particles is much smaller than $k_F^{-1}$, the many-body wave function takes a universal asymptotic form as a pairwise wavefunction describing the relative motion of two particles. $C_{\alpha\neq\beta}$ must be required to characterise the coherence of such pairwise wavefunction in the angular momentum space as analogous to the textbook example of a spin, whose transverse magnetisation needs to be measured for probing the spin coherence, regardless of the choice of quantisation axis.
In particular, $C_{\alpha\neq\beta}$ is crucial for a generic system at low temperatures whose total angular momentum is not conserved, due to either anisotropic external potentials  or anisotropic interactions. 
Second, any physical quantity determined by the asymptotic pairwise wavefunction depends on both $C_{\alpha\alpha}$ and $C_{\alpha\neq \beta}$, if it is not angular momentum selective. A prototypical example is the momentum distribution $n(\bf k)$ at large $k=|{\bf k}|$, which crucially replies on $C_{\alpha\neq \beta}$.
Third, $C_{\alpha\neq \beta}$ allows one to access many-body physics beyond the scope of diagonal terms $C_{\alpha\alpha}$.
For instance, $C_{\alpha\neq\beta}$ directly reflects the phase coherence between different order parameters of a superfluid with mixed partial waves, and thus allows one to trace macroscopic quantum phenomena from short-range correlations.

\begin{figure}
{\includegraphics[width=0.2\textwidth]{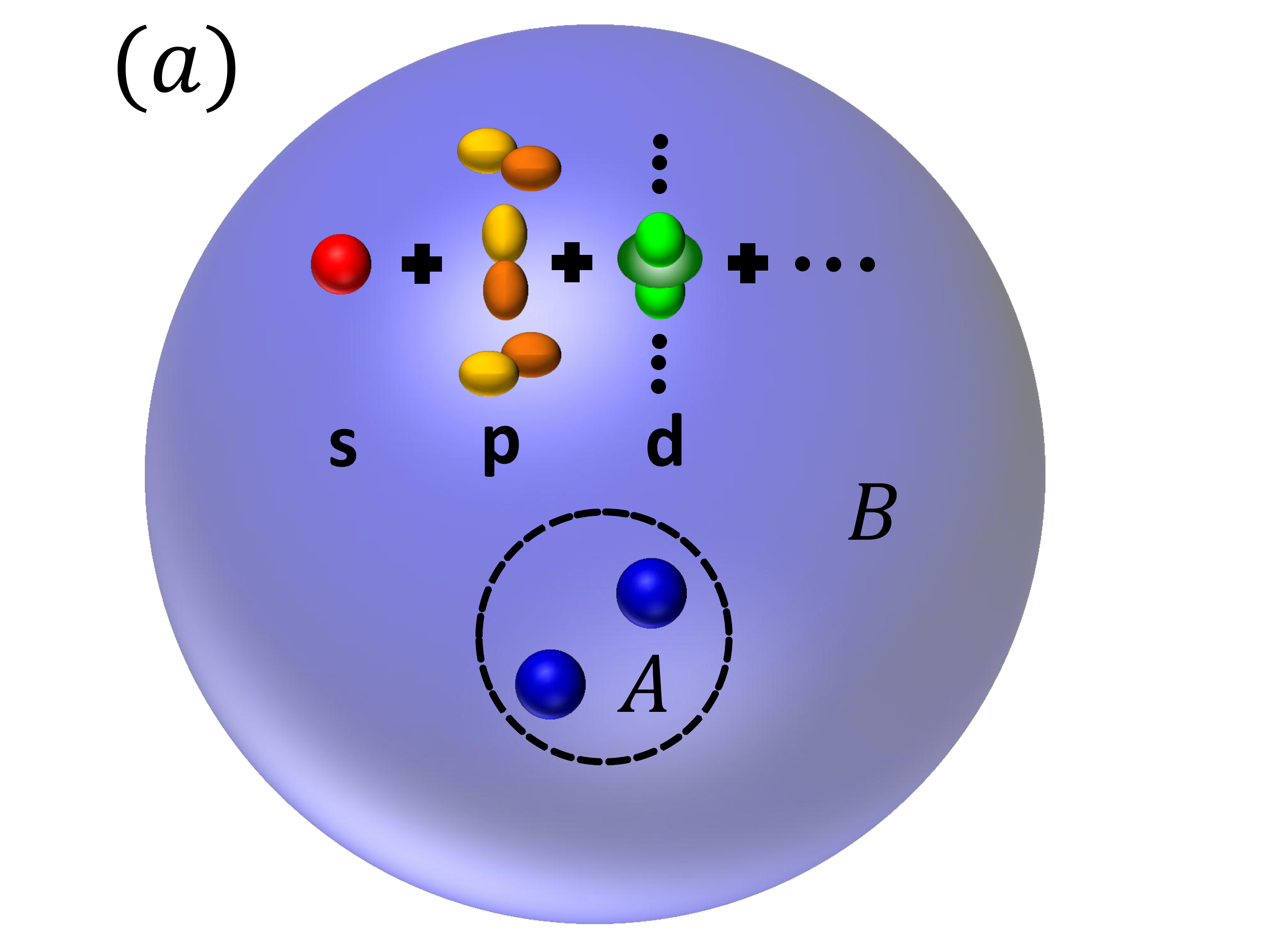}}
{\includegraphics[width=0.26 \textwidth]{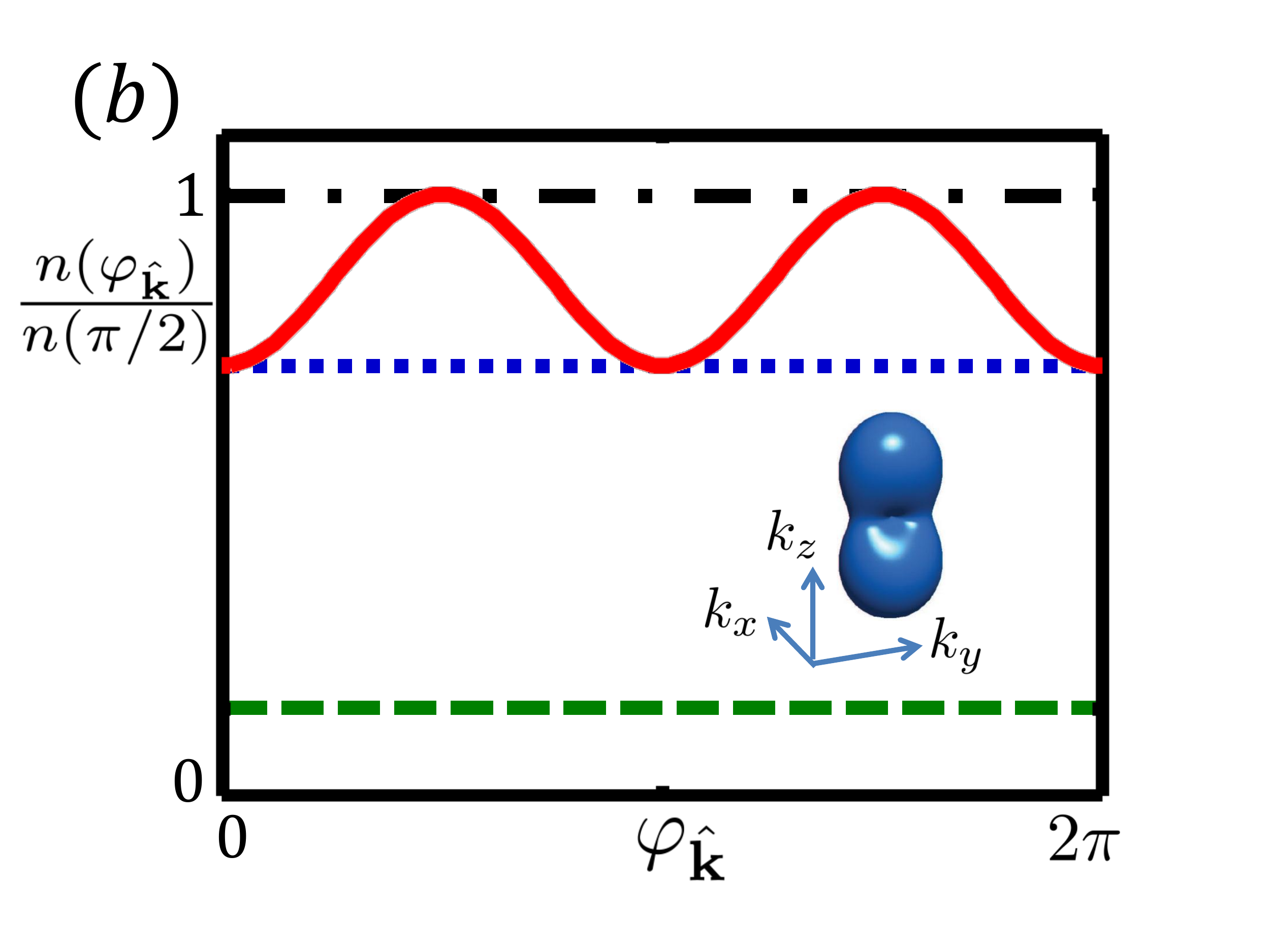}}
\caption{(a) small dark blue spheres represent two particles picked up from a dilute quantum system, which is represented by the big light blue cloud. The relative motion of such pair in general contains multiple partial waves and is entangled with the rest of the system. (b) the solid red curve represents the dependence of $n({\bf k})$ of an anisotropic $p$-wave superfluid $Y_{10}({\hat{\bf k}})+(\lambda/\sqrt{2})(Y_{11}({\hat{\bf k}})+Y_{1,-1}({\hat{\bf k}}))$ on the azimuthal angle $\varphi_{\hat{\bf k}}$ in the momentum space.  $\lambda=0.4$, $k/k_F\gg 1$, and the polar angle $\theta_{\hat{\bf k}}=\pi/4$. The blue dotted(green dashed) curve is the contribution from $C_{1010}$($C_{1111}$ or $C_{1-11-1}$), respectively. The black dash dotted curve includes the contribution from all diagonal contact.  Inset is the three-dimensional plot of $n({\bf k})$. The off-diagonal contact $C_{\alpha\neq\beta}$, which is proportional to $\Delta_{\alpha}^*\Delta_{\beta}$, gives rise to the difference between the solid and dash dotted curve. }
\end{figure}

To concretise discussions, we consider a single component system with $N$ particles, whose total angular momentum does not need to be a good quantum number. When the $i$th and the $j$th particles are close to each other, the asymptotic form of the many body wavefunction in three dimensions $\Psi({\bf r}_1,{\bf r }_2,...,{\bf r}_N)$ can be written as
\begin{equation}
\Psi ({\bf r}_1,{\bf r }_2,...,{\bf r}_N)  \stackrel{|{\bf r}_{ij}|\ll k_F^{-1}}{\xrightarrow{\hspace*{1cm}} } \sum_{\alpha} \psi_{\alpha}({\bf r}_{ij})G_{\alpha}({\bf R}_{ij}) \label{Asym}
\end{equation}
where $\alpha=(l, m)$ is the quantum number for the angular momentum in three dimensions, ${\bf r}_{ij}={\bf r}_i-{\bf r}_j$ is the real space coordinate of the relative motion of the $i$th and $j$th particle, ${\bf R}_{ij}=\{\frac{{\bf r}_i+{\bf r}_j}{2},{\bf r}_{k\neq i,j}\}$ is a short-hand notation, including the center of mass coordinate of $i$th and $j$th particles and the coordinates of all other ones.  $\psi_{\alpha}({\bf r}_{ij})$ is an unnormalized zero-energy solution of the Hamiltonian, $\hat{h}=-\frac{\hbar^2}{M}\nabla^2+\hat{U}({\bf r}_{ij})$. To simplify notations, we have considered the zero energy expansion of the two-body wave function $\psi_{\alpha}({\bf r}_{ij})\equiv\psi_{\alpha}({\bf r}_{ij};0)  \approx  \psi_{\alpha}({\bf r}_{ij};\epsilon)$, where $\epsilon$ is the energy of the relative motion of the $i$th and $j$th particles. Though the energy dependence of $\psi_{\alpha}({\bf r}_{ij};\epsilon)$ gives rise to interesting structures of contact of each partial wave scattering, it does not affect discussions of the main results regarding the contact matrix. Thus, in the main text, we focus on such zero energy expansion. Finite energy corrections are given in the supplementary material. 

We now consider a class of operators $\hat{O}$, which relies on the short-range behaviour of the relative motion of a pair of particles. The interaction energy $\hat{U}_{int}=\sum_{i<j} U({\bf r}_i-{\bf r}_j)$ is such example. Recall that we consider short-range interaction, the range  $r_0$ of which is much smaller than the interparticle spacing, one sees that the expectation value of $\hat{O}$ is indeed determined by the asymptotic form of the many-body wave function, as shown in Eq.(\ref{Asym}). Other well known examples in this category of operators include momentum distribution $n({\bf k})$ at large momentum $k$, photoassociation rate, rf-spectroscopy, and etc \cite{Tan1, Tan2, Tan3, V1,V2,Vale,T1,T2, rf1,rf2,rf3,rf4}. 

Using Eq.(\ref{Asym}), we obtain the expectation value of $\hat{O}$,
\begin{equation}
\langle \hat{O}\rangle=\sum_{\alpha\beta} \frac{1}{32\pi^2} \langle \psi_{\alpha}|\hat{O}|\psi_{\beta}\rangle  C_{\alpha\beta}, \label{O}
\end{equation}
where  $\langle \psi_{\alpha}|\hat{O}|\psi_{\beta}\rangle=\int d{\bf r}_{ij}  \psi_{\alpha}^*({\bf r}_{ij})\hat{O} \psi_{\beta}({\bf r}_{ij})\equiv O_{\alpha\beta}$ is a quantity purely determined by two-body physics, and
\begin{equation}
C_{\alpha\beta}=32\pi^2 \frac{N(N-1)}{2}  \int d{\bf R}_{ij} G^*_{\alpha}({\bf R}_{ij})G_{\beta}({\bf R}_{ij})\label{c}
\end{equation}
encodes all many-body physics. The prefactor $\frac{N(N-1)}{2}$ comes from the number of pairs of identical particles and $32\pi^2$ is introduced to simplify later expressions of $n({\bf k})$. When $\alpha=\beta$, Eq.(\ref{c}) recovers the contact we defined for an arbitrary partial wave scattering \cite{Zhou1}. If the total angular momentum $L$ is conserved,  $\alpha$ uniquely fixes  the angular momenta of both  the pair of particle and the rest of the system, which is represented by ${\bf R}_{ij}$, so that $G_{\alpha}({\bf R}_{ij})$ must be orthogonal to each other. However, in a generic system with broken rotational symmetry, due to either anisotropic external trapping potential or anisotropic interaction,  different $G_{\alpha}({\bf R}_{ij})$ may not be orthogonal to each other, i.e., $\int d{\bf R}_{ij} G^*_{\alpha}({\bf R}_{ij})G_{\beta}({\bf R}_{ij})\neq 0$ when $\alpha\neq\beta$. Thus the off-diagonal contact $C_{\alpha\neq\beta}$ becomes finite.

Eq.(\ref{Asym}) and Eq.(\ref{O}) allow one to fully unveils the structure of the pairwise correlations at short distance.  Formally, it is equivalent to a bipartite decomposition of an arbitrary many-body system into two parts. The relative motion of an arbitrarily picked up pair of particles is regarded as one (small) subsystem $A$, and the rest of the many-body system, including the center of mass of such pair and all other $N-2$ particles, is regarded as the other (big) subsystem $B$, as shown in Fig. 1(a). Since $\hat{O}$ only acts on the subsystem $A$, its expectation value also depends on the overlap integral of the wavefunctions of subsystem $B$. The off-diagonal terms $ C_{\alpha\neq\beta}$ thus characterises the coherence of the subsystem $A$ in  the angular momentum space and how much it is entangled with the subsystem $B$, which can be viewed as the environment of $A$.  Two extreme cases can be used to illuminate the physics. 
\begin{eqnarray}
\int d{\bf R}_{ij} G^*_{\alpha}({\bf R}_{ij})G_{\beta}({\bf R}_{ij})=0\,\,\,(\alpha\neq\beta), & \mathrm{case (I)} \\ \label{cs1}
\forall\alpha,   G_{\alpha}({\bf R}_{ij})=G({\bf R}_{ij}), \,\,\,\,\,\,\,\,\, & \mathrm{case (II)} \label{cs2}
\end{eqnarray}
In case (I),  the off-diagonal contact $C_{\alpha\neq\beta}$ vanishes, and each $\psi_{\alpha}({\bf r}_{ij})$ is coupled to a unique one in an orthogonal set of wavefunctions $G_{\alpha}({\bf R}_{ij})$. In other words, the relative motion of the pair is highly entangled with the rest of the system, and thus losses its own coherence. In case (II), all $G_{\alpha}({\bf R}_{ij})$ are identical, and Eq.(\ref{cs2}) is satisfied for any $\alpha$. The right hand side of Eq.(\ref{Asym}) becomes $[\sum_{\alpha} \psi_{\alpha}({\bf r}_{ij})]G({\bf R}_{ij})$, which is a product state. In this case, the relative motion of the pair is not entangled with the rest of the system at all, and its own coherence retains. Later, we will discuss examples of many-body states in both cases, i.e., quantum Hall states in (I) and BCS-superfluids in (II), respectively. 

It is worth mentioning that Eq.(\ref{O}) is analogous to the central spin problem, in which the coherence of the electronic spin is controlled by the entanglement with a bath of nuclear spins, i.e., $\langle\hat{\sigma}_x \rangle_e\sim  \langle I_1|I_2\rangle$ for an entangled state $|\uparrow\rangle_e|I_1\rangle+|\downarrow\rangle_e|I_2\rangle$, where $|I_1\rangle$($|\uparrow\rangle_e$) and $|I_2\rangle$($|\downarrow\rangle_e$) are nuclear(electronic) spin states\cite{Milburn, Sham1, Sarma, Sham2}. Spin coherence allows one to detect a wide range of many-body physics in the bath\cite{Sun, Suter, Liu1,Liu2,Liu3}. Here, $\alpha$ may be regarded as a pseudospin index. Moreover, a unique feature is that the $A$ subsystem is actually a pair of particles within the many-body system of interest. Due to the length scale separation $k_Fr_0\ll 1$, and the resultant Eq.(\ref{Asym}) and Eq.(\ref{O}), many thermodynamic quantities of the many-body system, such as the momentum distribution $n({\bf k})$, depends on the off-diagonal contact $C_{\alpha\neq\beta}$. More importantly, Eq.(\ref{O}) allows one to trace many-body physics using observable that are dependent on the pairwise correlations at short distance. For instance, the symmetry and the coherence of the order parameters in a superfluid can be traced from the momentum distributions, as shown later. 

The momentum distribution,  a typical measurable in ultracold atoms,  can be computed using $n_{\bf k}=\sum_{i}\int \prod_{j\neq i}d{\bf r}_{j}\Big|\int d{\bf r}_i\Psi({\bf r}_1,{\bf r}_2,... {\bf r}_N)e^{ -i {{\bf k}\cdot {\bf r}_i}}\Big|^2$. It has been shown that for large $k\gg k_F$, the form of $n({\bf k})$ is determined by the Fourier transform of the asymptotic form in Eq.(\ref{Asym}), i.e., $\psi_{\bf k}({\bf R}_{ij})\equiv \int d{\bf r}_{ij} e^{ -i {\bf k}\cdot {\bf r}_{ij}}[\sum_{\alpha} \psi_{\alpha}({\bf r}_{ij})G_{\alpha}({\bf R}_{ij}) ] $. Recall that $\psi_\alpha({\bf r}_{ij})=\varphi_{\alpha}(r_{ij})Y_{\alpha}(\hat{\bf r}_{ij})$, where $\varphi_{\alpha}(r_{ij})$ is the radial part of the wavefunction, $r_{ij}=|{\bf r}_{ij}|$, $\hat{\bf r}_{ij}={\bf r}_{ij}/| {\bf r}_{ij}|$, and $Y_{\alpha}(\hat{\bf r}_{ij})$ is the spherical Harmonics with $\alpha=(l,m)$, and that $e^{i{\bf k}\cdot{\bf r}}=4\pi\sum_{l=0}^{\infty}\sum_{m=-l}^{l}i^lj_l(kr)Y^*_{lm}(\hat{\bf k})Y_{lm}(\hat{\bf r})$, where $j_l(kr)$ is the first kind spherical Bessel function, one sees that $\psi_{\bf k}({\bf R}_{ij})$ is a superposition of many partial waves in the momentum space. $|\psi_{\bf k}({\bf R}_{ij})|^2$ thus naturally has the cross terms $G^*_\alpha({\bf R}_{ij})G_\beta({\bf R}_{ij})$.  Consider a short-range interaction, in the regime $k_F\ll k\ll r_0^{-1}$, $\varphi_\alpha(r_{ij})$ has the asymptotic form $r^{-l-1}_{ij}$. A straightforward calculation shows that 
\begin{equation}
\label{nk}
\begin{split}
n_{\bf k} &\stackrel{k_F\ll k\ll r_0^{-1}} {\xrightarrow{\hspace*{1cm}} } \sum \limits_{lm} { \frac{ C_{lmlm}}{k^{4-2l}} |Y_{lm}(\hat {\bf k})|^2 } \\
& + \sum \limits_{(l,m) \neq (l',m')} {i^{l-l'} \frac{C_{lml'm'}}{k^{4-l-l'} } Y^*_{lm}(\hat {\bf k}) Y_{l'm'}(\hat {\bf k}) } 
\end{split} 
\end{equation}
Details of the calculation are presented in the supplementary materials. 

Eq.(\ref{nk}) readily allows one to measure contact matrix in experiments by fitting the angular part of the momentum distribution based on the partial wave expansion in this equation. The first line in Eq.(\ref{nk}) is the contribution from each individual contact, which has been discussed before. The second line comes from the off-diagonal contact. It inevitably leads to not only new power-law dependence $k^{l+l'-4}$, but also interference pattern in the momentum space. Such interference pattern is a direct probe of the coherence of the pairwise wave function in Eq.(\ref{Asym}) in the angular momentum space. For many-body states in case (I), such interference pattern vanishes. In contrast, the amplitude of the interference pattern of many-body states in case (II) directly reflects the strength of the off-diagonal contact $C_{\alpha\neq\beta}$.

Whereas we have been focusing on three dimensions, it is rather clear that the above discussions can be directly generalised to two dimensions. One defines contact matrix in two dimensions,
\begin{equation}
C_{\alpha\beta}=8\pi^2 \frac{N(N-1)}{2}  \int d{\bf R}_{ij} G^*_{\alpha}({\bf R}_{ij})G_{\beta}({\bf R}_{ij}).
\end{equation}
The only difference from Eq.(\ref{c}) is the prefactor. Here $\alpha=l$ represents the angular momentum quantum number.  We do not use a different symbol to denote contact in two dimensions, since it is rather apparent in later discussions whether it means the one in three dimensions or two dimensions. The expression for the momentum distribution in Eq.(\ref{nk}) remains unchanged. 

We now discuss how to use contact matrix to trace many-body physics and macroscopic quantum phenomena. In a recent work of us, we have discussed contact of quantum Hall states, which belong to case (I). In such states, the off-diagonal contact vanishes, and each pair of particles is highly entangled with the rest of the system. We have used all the diagonal contact $C_{\alpha}$ to define contact spectrum $\{C_{\alpha}\}$, which serves as a unique tool to probe the quantum Hall states. Here, we focus on BCS-superfluids, which belong to case (II).  

It is well known that the first quantisation form of a BCS wavefunction of a (spinless) superfluid is written as
\begin{equation}
\Psi_{BCS}=\mathcal{A}[ \phi({\bf r}_1-{\bf r}_{2})\phi({\bf r}_3-{\bf r}_{4})\cdots \phi({\bf r}_{N-1}-{\bf r}_{N})]
\end{equation}
where $\mathcal{A}$ is the antisymmetrizing operator, $\phi({\bf r}_i-{\bf r}_j)$ is the pair wavefunction.  Considering the asymptotic behaviour when ${\bf r}_i\rightarrow{\bf r}_j$, one sees that it is indeed described by Eq.(\ref{cs2}). This is not surprising, since a BCS wave function can be viewed as a condensate of pairs, in which each pair is not entangled with others.  Sometimes, $\phi({\bf r}_i-{\bf r}_j)$ contains only a single partial wave $Y_{\alpha}(\hat{\bf r}_{ij})$. Nevertheless, many important superfluids are mixtures of multiple partial waves. One example is the cyclic state of $l=2$ superfluid. The order parameter $\Delta_{\bf k}$ can be written as $\Delta_{\bf k}=\Delta(k) \tilde \Delta(\hat {\bf k})$, $\tilde \Delta(\hat {\bf k})=e^{-\pi i/6}/2(Y_{22}(\hat{\bf k}) +Y_{2-2}(\hat{\bf k}))+e^{4\pi i/3}/\sqrt{2} Y_{20}(\hat{\bf k}) \propto \hat k_x^2+e^{2\pi i/3}\hat k_y^2+e^{4\pi i/3}\hat k_z^2$  \cite{Mermin}. It is a superposition of different magnetic quantum numbers $m$.  The other example is the anisotropic $p$-wave superfluid, when the an anisotropic interaction, such as dipole-dipole interaction, breaks the rotation symmetry \cite{Yip}. In two dimensions, $s+d$ superfluid is an important example for discussing topological phase transitions. It is a mixture of $l=0$ and $l=2$. These superfluids have a unique feature that the order parameter is a coherent superposition of different partial wave components, distinct from incoherent mixtures of multiple order parameters. However, it remains challenging to directly probe this phase coherence.  We discuss how to use contact matrix and the momentum distributions to access such phase coherence.

To compute the contact matrix of a superfluid, it is simple to make use of the second quantisation form to obtain the momentum distribution. Consider a single component Fermi gases with short range interaction $V({\bf r})$, the Hamiltonian $\hat H=\hat H_0 +\hat V$, $\hat H_0=\sum \nolimits_{\bf k} \epsilon_{\bf k} \hat{a}^\dag_{{\bf k}}\hat{a}_{{\bf k}}$, $\epsilon_{\bf k}=\hbar^2 k^2/(2M)$, $M$ is the mass of each particle, 
\begin{equation}
\hat V=\Omega^{-1}  \sum \nolimits_{{\bf k},{\bf k'}} V_{{\bf k'}-{\bf k}} \hat{a}^\dag_{{\bf k'}}\hat{a}^\dag_{-{\bf k'}}\hat{a}_{-{\bf k}}\hat{a}_{{\bf k}},
\end{equation}
$\Omega$ is the volume of the system, $V_{{\bf k'}-{\bf k}}=\int d {\bf r} e^{-i({\bf k'}-{\bf k})\cdot {\bf r}} V({\bf r})$ is the Fourier transform of the interaction of $V({\bf r})$.  In standard BCS theory, $\left| {G} \right\rangle =\prod\nolimits_{\bf k} (u_{\bf k}+v_{\bf k} \hat{a}_{{\bf k}}^\dag \hat{a}_{-{\bf k}}^\dag)\left| {0} \right\rangle$, $|u_{\bf k}|^2+|v_{\bf k}|^2=1$, and the momentum distribution is written as  $n({\bf k})= \left( 1-(\epsilon_{\bf k}-\mu)/{E_{\bf k}} \right)/2$, where $E_{\bf k}=\sqrt{(\epsilon_{\bf k}-\mu)^2+|\Delta_{\bf k}|^2}$ and $\mu$ is the chemical potential. Note that $\Delta_{\bf k}$ in general may contain multiple partial waves, 
 \begin{equation}
 \Delta_{\bf k}=\sum \limits_{lm} (-i)^l\Delta_{lm} k^l Y_{lm}(\hat {\bf k}),\label{ge}
 \end{equation}
 where $\Delta_{lm}$ is the strength of the order parameter in a given partial wave channel. Eq.(\ref{ge}) is valid for $k\ll k_{l}$, where $k_{l}\sim 1/r_0$ is a momentum cutoff that reproduces the realistic two-body scattering phase shift \cite{schimtt,Ho}. 
 
Under the condition $\epsilon_{\bf k} \gg \mu$ and $\epsilon_{\bf k} \gg |\Delta_{\bf k}|^2$, one obtains the momentum distribution at large $k$,
\begin{equation}
\label{snk}
n({\bf k}) \stackrel{k_F\ll k\ll r_0^{-1}} {\xrightarrow{\hspace*{1cm}} }\sum\limits_{lml'm'} i^{l-l'}\frac{M^2}{ \hbar^4}\frac{ \Delta_{lm}^* \Delta_{l'm'}}{k^{4-l-l'}} Y^*_{lm}(\hat {\bf k}) Y_{l'm'}(\hat {\bf k})
\end{equation} 
Compare it with Eq.(\ref{nk}), one sees that in such BCS superfluid, contact matrix is directly related to superfluid order parameters, 
\begin{equation}
\label{gap}
C_{lml'm'}= \frac{M^2}{ \hbar^4} \Delta_{lm}^* \Delta_{l'm'} 
\end{equation}  
Eq.(\ref{gap}) thus establish a direct relation between contact matrix and the superfluid order parameters. In particular, the phase coherence between different order parameters is revealed by the off diagonal contact. As a demonstration, Fig.1(b) shows the momentum distribution of an anisotropic $p$-wave superfluid $\hat{k}_z-i\lambda \hat{k}_y\propto Y_{10}({\hat{\bf k}})+(\lambda/\sqrt{2})(Y_{11}({\hat{\bf k}})+Y_{1,-1}({\hat{\bf k}}))$. For such superfluid, the contact matrix is a $3\times 3$ one. The off-diagonal contact $C_{\alpha\neq\beta}$, such as $C_{101-1}$, $C_{1011}$, and $C_{111-1}$, which is proportional $\Delta_{\alpha}^*\Delta_{\beta}$, gives rise to the dependence of $n({\bf k})$ on the azimuthal angle $\varphi_{\hat{\bf k}}$ in the momentum space, i.e., the difference between the red solid curve and the black dash dotted one in Fig.1(b).

Besides momentum distributions, it is useful to comment on the relations between contact matrix and other universal relations. S. Tan first showed that the internal energy of a dilute system with $s$-wave scattering can be written as a functional of the momentum distribution\cite{Tan1}. In a single component system, such integral is written as $ E=\int \frac{d{\bf k}}{(2\pi)^3} \frac{\hbar^2 k^2}{2M}(n({\bf k}) -C_{0000}/k^4 |Y_{00}(\hat {\bf k})|^2) +\frac{\hbar^2C_{0000}}{32\pi^2 Ma_0}$ \cite{V4}. Such a functional fixes the divergent problem in the zero-range interaction limit.  In a recent work, we generalise such functional to a generic short range interaction, which include contact of all partial wave scatterings, i.e., the  diagonal terms $C_{\alpha\alpha}$ in contact matrix \cite{Zhou1}.  Here, we have verified that such energy functional is not affected by the off-diagonal terms. Apparently, the cross terms $C_{\alpha \neq \beta}$ in Eq.(\ref{snk}) do not contribute to the integral $\int d{\bf k} k^2n({\bf k})$, due to the orthogonal condition $\int d\Omega_{\bf k} Y^*_{\alpha}(\hat{\bf k})Y_{\beta}(\hat{\bf k})=\delta_{\alpha\beta}$, where $\int d\Omega_{\bf k} $ denotes the angular part of the integral in the momentum space.  S. Tan also found out the adiabatic relation, in which the derivative of the energy with respect to the inverse of the $s$-wave scattering length is given by $s$-wave contact, i.e., $\frac{dE}{d(-1/a_0)}\sim C_{0000}$\cite{Tan2}. Such relation were recently generalised to high partial wave scatterings\cite{P1,P2,P3,P4}. Here, the off-diagonal contact $C_{\alpha\beta}$ is apparently not associated with any scattering length and thus is beyond the scope of the conventional adiabatic relations. Results of photoassociation and rf-spectroscopy are presented in the supplementary material.

It is useful to highlight a few potential applications of contact matrix. Anisotropic interactions, such as the magnetic dipole-dipole interaction, are  important in many cases, for instance, the $p$-wave Feshbach resonance \cite{Jin, S1,S2}. Recent experiments on Er and Dr has also shown the importance of anisotropic interaction in magnetic lanthanide atoms \cite{L1,L2}. For those anisotropic interactions, different partial waves naturally mix with each other in two-body scattering. A full description of contact matrix is thus required and our general results apply.

We have shown that contact matrix provides one a complete description of the  pairwise correlations at short distance in a dilute quantum system. The off-diagonal contact, which elude previous studies, determines thermodynamic quantities and allows one to trace macroscopic quantum phenomena. Whereas we focus on single component systems here, it is straightforward to generalise our results to multi-component systems. We hope that our work will inspire more interests in applying contact matrix in many-body physics.

This work is supported by RGC/GRF(14306714). 

*SZ and MH contribute equally to this work.

\vspace{0.2in}

\onecolumngrid

\vspace{0.2in}

\centerline{\bf Supplementary Materials}

\vspace{0.1in}

In this supplementary material, we present the results on the large momentum distribution including the finite energy corrections, photoassociation and rf-spectroscopy.

\vspace{0.15in}

{\bf Large momentum distribution and contact matrix}

For a $N$-particle dilute quantum systems with short range interaction, the many-body wave function ${\Psi } = \Psi \left( {{{\bf r}_1},{{\bf r}_2}, \cdots ,{{\bf r}_N}} \right)$, asymptotically, can be written as
\begin{equation}
\Psi \stackrel{r_{ij}\ll k_F^{-1}}{\xrightarrow{\hspace*{1cm}} } \sum_{lm} \int d\epsilon  \psi_{lm}({\bf r}_{ij};\epsilon)G_{lm}({\bf R}_{ij};E-\epsilon)
\end{equation}
where ${\bf r}_{ij} = {\bf r}_i -{\bf r}_j$, $ r_{ij}=|{\bf r}_{ij}|$, ${\bf R}_{ij}=\{({\bf r}_i + {\bf r}_j)/2, {\bf r}_{k \neq i,j}\} $, $\epsilon =q^2_{\epsilon}/M$ is the energy of the the relative motion of the $i$th and $j$th particle pair, and
\begin{equation}
{\psi _{lm}}\left( {{{\bf r}_{ij}};\epsilon } \right) = \frac{q_{\epsilon}^{l+1}}{\tan{[\eta_l]}}\left\{ {{j_l}\left( {{q_\epsilon }{r_{ij}}} \right) - \tan \left[ {{\eta _l}} \right]{n_l}\left( {{q_\epsilon }{r_{ij}}} \right)} \right\}{Y_{lm}}(\hat{\bf r}_{ij})
\end{equation}
is the solution of the relative motion of an isolated two-body system. The momentum distribution is computed using the expression in the main text $
{n_{\bf k}} = \sum\limits_i {\int {\prod\nolimits_{j \ne i} {d{{\bf r}_j}} {{\left| {\int {d{{\bf r}_i}\Psi \left( {{{\bf r}_1},{{\bf r}_2}, \cdots ,{{\bf r}_N}} \right){e^{ - i {\bf k} \cdot {{\bf r}_i}}}} } \right|}^2}} }$. By defining $ {\Psi _i}\left( {\bf k} \right) = \int {d{{\bf r}_i}\Psi \left( {{{\bf r}_1},{{\bf r}_2}, \cdots ,{{\bf r}_N}} \right){e^{ - i {\bf k} \cdot {{\bf r}_i}}}} $, we obtain ${n_{\bf k}} = \sum\limits_i {\int {\prod\nolimits_{j \ne i} {d{{\bf r}_j}} {{\left| {{\Psi _i}\left( {\bf k} \right)} \right|}^2}} }.$
By defining a pure two-body quantity, ${F_{lm}}\left( {{\bf k};\epsilon } \right) = \int {d{{\bf r}_{ij}}{\psi _{lm}}\left( {{{\bf r}_{ij}};\epsilon } \right){e^{ - i {\bf k} \cdot \left( {{{\bf r}_i} - {{\bf r}_j}} \right)}}}$
so that, in the regime $|{\bf k}| \gg k_F$, one has
\begin{equation}
{\Psi _i}\left( {\bf k} \right) \to \sum\limits_{j \ne i} {{e^{ - i{\bf k} \cdot {{\bf r}_j}}}} \sum\limits_{lm} {\int {d\epsilon {G_{lm}}\left( {{{\bf R}_{ij}};E - \epsilon } \right){F_{lm}}\left( {{\bf k};\epsilon } \right)} }.
\end{equation}
If one extends the wave function $\psi_{lm}({\bf r}_{ij};\epsilon)$ in the regime $[r_0, \infty]$ to $[0,\infty]$, $F_{lm}({\bf k}; \epsilon)$ in the regime $k_F \ll k$, $k=|{\bf k}|$, can be given by
\begin{equation}
\begin{split}
{F_{lm}}\left( {{\bf k};\epsilon } \right) &= \int {d{{\bf r}_{ij}}\left( {\frac{{{\beta _{l0}}}}{{r_{ij}^{l + 1}}} + \frac{{{\beta _{l1}}q_\epsilon ^2}}{{r_{ij}^{l - 1}}} +  \cdots } \right){Y_{lm}}\left( \hat{\bf r}_{ij} \right){e^{ - i{\bf k} \cdot {{\bf r}_{ij}}}}} \\
 &= 4\pi {\left( { - i} \right)^l}\left( {{k^{l - 2}} + q_\epsilon ^2{k^{l - 4}} +  \cdots } \right){Y_{lm}}( {\hat{\bf k}} )
\end{split}
\end{equation}
where one has used $\beta_{ls}=(2l-2s-1)!!/(2s)!!$, ${e^{ - i{\bf k} \cdot {{\bf r}_{ij}}}} = 4\pi \sum\nolimits_{l = 0}^\infty  {\sum\nolimits_{m =  - l}^l {{{\left( { - i} \right)}^l}{j_l}\left( {k{r_{ij}}} \right){Y_{lm}}( \hat{\bf k} )Y_{lm}^ * ( \hat{\bf r}_{ij} )} }$ and $f_{ls}(k \Lambda)=\int_0^{k\Lambda} {\left[ {{\beta _{ls}}/{x^{l - 2s - 1}}} \right]{j_l}\left( x \right)dx}  = 1 - {\beta _{ls}}\sum\nolimits_{i = 0}^s {\left[ {\left( {2s} \right)!!{j_{l - i - 1}}\left( {k\Lambda } \right)} \right]/[(2s-2i)!!{{\left( {k\Lambda } \right)}^{l - 2s + i - 1}}]}  \stackrel{k \Lambda \to \infty}{\xrightarrow{\hspace*{1cm}} }  1$. One can then obtain
\begin{equation}
{\Psi _i}\left( {\bf k} \right) \to \sum\limits_{j \ne i} {{e^{ - i{\bf k} \cdot {{\bf r}_j}}}} \sum\limits_{lm} {4\pi {{\left( { - i} \right)}^l}\left( {g_{lm}^0{k^{l - 2}} + g_{lm}^2{k^{l - 4}} +  \cdots } \right){Y_{lm}}( \hat{\bf k} )},
\end{equation}
where $g_{lm}^{s} = \int {d\epsilon q_\epsilon ^{s}{G_{lm}}\left( {{{\bf R}_{ij}};E - \epsilon } \right)}$. And since the cross term $e^{i{\bf k} \cdot ({\bf r}_{j} - {\bf r}_{j'} )}$ vanishes in the large $k$ limit, and $d{{\bf R}_{ij}}  = \prod\nolimits_{k \ne i,j} {d{{\bf r}_k}} d\left( {{{\bf r}_i} + {{\bf r}_j}} \right)/2$, one has
\begin{equation}
{n_{\bf k}} \stackrel{k_F\ll k\ll r_0^{-1}} {\xrightarrow{\hspace*{1cm}} } \sum\limits_{lml'm'} {{i^{l - l'}}\left( {{k^{l+l' - 4}}{C_{lml'm'}} + {k^{l+l' - 6}}C_{lml'm'}^1 + {k^{l+l' - 8}}C_{lml'm'}^2 +  \cdots } \right)Y_{lm}^ * ( \hat{\bf k}){Y_{l'm'}}( \hat{\bf k} )}
\end{equation}
which mainly includes all the terms that can cause the divergence in the energy functional discussed in reference\cite{Zhou}, i.e., $k^{s \geq-4}$, where we define a contact matrix $\{ C_{lml'm'} \}$ as
\begin{equation}
C_{lml'm'}^{ij} = {\left( {4\pi } \right)^2}N\left( {N - 1} \right)\int {d{{\bf R}_{ij}}g_{lm}^{2i * }g_{l'm'}^{2j}} ,\quad  C_{lml'm'} =C_{lml'm'}^{00} , \quad C_{lml'm'}^{s>0} = \sum\limits_{j = 0}^s {C_{lml'm'}^{\left( j \right)\left( {s - j} \right)}}.
\end{equation}

Note that for the 2D dilute quantum systems, one also has
\begin{equation}
{\psi _{l}}\left( {{{\bf r}_{ij}};\epsilon } \right) =\frac{\pi}{2} \frac{q_{\epsilon}^{l}}{\tan{[\eta_l]}}\left\{ {{J_l}\left( {{q_\epsilon }{r_{ij}}} \right) - \tan \left[ {{\eta _l}} \right]{N_l}\left( {{q_\epsilon }{r_{ij}}} \right)} \right\}{Y_{l}}(\hat{\bf r}_{ij})
\end{equation}
where $Y_l( \hat{\bf r}_{ij})=e^{il \theta_{ij}}/\sqrt{2\pi}$, $\theta_{ij}=\mathrm{arg}\{{\bf r}_{ij}\}$. By repeating the above steps, one end up with the same results.
\\

\vspace{0.15in}

{\bf Photoassociation}

Photoassocation couples two atoms to an electronically excited molecular state. If the photoassociation is not angluar momentum selective, multiple partial waves can all be coupled to the excited state. For a two-body problem, the coupling matrix element of each partial wave to the excited state is written as 
\begin{equation}
V^{[2]}_\alpha=\sqrt{\frac{2\pi I}{c}}\langle \varphi_e|{\bf d}_M\cdot\hat{\bf e}|\psi_\alpha\rangle=\sqrt{\frac{2\pi I}{c}}\Omega_\alpha\big|{\bf d}_M\big|
\end{equation}
where $c$ is the speed of light, $I$ is the laser intensity, $\varphi_e$ is the electronically excited molecule wave function, $\psi_\alpha$ is the two-body wave function of $\alpha$-th partial wave, ${\bf d}_M$ is the electric dipole moment, $\hat{\bf e}$ is the polarization vector of the laser and $\Omega_\alpha=\langle \varphi_e|\hat{\bf d}_M\cdot\hat{\bf e}|\psi_\alpha\rangle$ where $\hat{\bf d}_M={\bf d}_M/\big|{\bf d}_M\big|$\cite{PA1,PA2,PA3}.

For a many-body problem, since the molecule size is much smaller than the average interparticle distance, $G_\alpha({\bf R}_{ij})$ can be considered as a normalization factor for $\psi_\alpha({\bf r}_{ij})$, the coupling strength is written as
\begin{equation}
\begin{split}
\Gamma&=\frac{2\pi I}{c}\frac{N(N-1)}{2}\int d{\bf R}_{ij}\big|\sum_\alpha\langle \varphi_e|{\bf d}_M\cdot\hat{\bf e}|\psi_\alpha\rangle G_\alpha({\bf R}_{ij})\big|^2=\frac{I}{16\pi c}\big|{\bf d}_M\big|^2\sum_{\alpha,\beta}\Omega^*_\alpha\Omega_\beta C_{\alpha\beta}
\end{split}
\end{equation}
The off-diagonal contact thus determines $\Gamma$. 

\vspace{0.15in}

{\bf Radiofrequency spectroscopy}

A rf field transfers atoms to an different hyperfine spin state. When the rf frequency between two hyperfine spin states $\omega$ is large compared with relevant frequencies in many-body physics, the transition rate has the asymptotic form\cite{SR}
\begin{equation}
\Gamma_\mathrm{rf}=2\pi\Omega^2_\mathrm{rf}\hbar\sum_{\bf k}{n}_{\bf k}\delta(\hbar\omega-\hbar^2k^2/M)
\end{equation}
where $\Omega_\mathrm{rf}$ is the rf Rabi frequency. Using the result of momentum distribution in equation (6) of the supplementary materials, one obtain
\begin{equation}
\begin{split}
\Gamma_\mathrm{rf}&=2\pi\Omega^2_\mathrm{rf}\hbar\frac{\Omega}{(2\pi)^3}\int d{\bf k}{n}_{\bf k}\delta(\hbar\omega-\hbar^2k^2/M) \\
&\approx\frac{\Omega^2_\mathrm{rf}M}{8\pi^2\hbar}\Omega\sum_{lm}\Big\{C_{lmlm}\Big(\frac{M\omega}{\hbar}\Big)^{l-3/2}+C^1_{lmlm}\Big(\frac{M\omega}{\hbar}\Big)^{l-5/2}+\cdots\Big\}
\end{split}
\end{equation}
where $\Omega$ is the volume of the system, $C_{lmlm}^{s\ge 1}$ are contact we defined for a high partial wave scattering associated with the subleading divergent terms and beyond in energy functional\cite{Zhou}. The off-diagonal contact does not contribute to rf spectroscopy, due to the orthogonality $\int d\Omega_{\bf k} Y^*_{\alpha}(\hat{\bf k})Y_{\beta}(\hat{\bf k})=\delta_{\alpha\beta}$, where $\int d\Omega_{\bf k} $ denotes the angular part of the integral in the momentum space.

\end{document}